\documentclass[conference]{IEEEtran}
\IEEEoverridecommandlockouts
\usepackage{cite}
\usepackage{amsmath,amssymb,amsfonts}
\usepackage{algorithmic}
\usepackage{graphicx}
\DeclareGraphicsExtensions{.pdf,.png,.jpg,.mps,.eps,.ps}
\graphicspath{{figs/}}
\usepackage{textcomp}
\usepackage{siunitx} 
\usepackage{xcolor}
\definecolor{mpcolor}{rgb}{1, 0.1, 0.59}
\definecolor{mdcolor}{rgb}{0.7, 0.3, 0.4}

\newenvironment{noticebox}[1][b]{
    \enlargethispage{3\baselineskip}
    \begin{figure}[#1]
    \footnotesize
}{
    \end{figure}
}
\def\BibTeX{{\rm B\kern-.05em{\sc i\kern-.025em b}\kern-.08em
    T\kern-.1667em\lower.7ex\hbox{E}\kern-.125emX}}
\begin{document}

\title{Real-Time Machine Learning Strategies\\for a New Kind of Neuroscience Experiments}
\author{
\IEEEauthorblockN{
    Ayesha Vermani%
    \IEEEauthorrefmark{1},
    Matthew Dowling%
    \IEEEauthorrefmark{1},
    Hyungju Jeon%
    \IEEEauthorrefmark{1},
    Ian Jordan%
    \IEEEauthorrefmark{2},
    Josue Nassar%
    \IEEEauthorrefmark{2},
    \\
    Yves Bernaerts%
    \IEEEauthorrefmark{1},
    Yuan Zhao%
    \IEEEauthorrefmark{3},
    Steven Van Vaerenbergh%
    \IEEEauthorrefmark{4},
    Il Memming Park%
    \IEEEauthorrefmark{1}\IEEEauthorrefmark{2}
    \thanks{This work was supported in part by NIH RF1 DA056404, Portuguese Recovery and Resilience Plan (PPR) 62, and Funda\c{c}\~{a}o para a Ci\^{e}ncia e a Tecnologia (FCT) project UIDB/04443/2020.}
}
\IEEEauthorblockA{\IEEEauthorrefmark{1}\textit{Champalimaud Centre for the Unknown}, Champalimaud Foundation, Lisbon, Portugal}
\IEEEauthorblockA{\IEEEauthorrefmark{2}\textit{RyvivyR}, USA \& Portugal}
\IEEEauthorblockA{\IEEEauthorrefmark{3}\textit{National Institute of Mental Health}, Bethesda, Maryland, USA}
\IEEEauthorblockA{\IEEEauthorrefmark{4}\textit{University of Cantabria}, Spain}
}

\maketitle

\begin{abstract}
Function and dysfunctions of neural systems are tied to the temporal evolution of neural states.
The current limitations in showing their causal role stem largely from the absence of tools capable of probing the brain's internal state in real-time.
This gap restricts the scope of experiments vital for advancing both fundamental and clinical neuroscience.
Recent advances in real-time machine learning technologies, particularly in analyzing neural time series as nonlinear stochastic dynamical systems, are beginning to bridge this gap.
These technologies enable immediate interpretation of and interaction with neural systems, offering new insights into neural computation.
However, several significant challenges remain.
Issues such as slow convergence rates, high-dimensional data complexities, structured noise, non-identifiability, and a general lack of inductive biases tailored for neural dynamics are key hurdles.
Overcoming these challenges is crucial for the full realization of real-time neural data analysis for the causal investigation of neural computation and advanced perturbation based brain machine interfaces.
In this paper, we provide a comprehensive perspective on the current state of the field, focusing on these persistent issues and outlining potential paths forward.
We emphasize the importance of large-scale integrative neuroscience initiatives and the role of meta-learning in overcoming these challenges.
These approaches represent promising research directions that could redefine the landscape of neuroscience experiments and brain-machine interfaces, facilitating breakthroughs in understanding brain function, and treatment of neurological disorders.
\end{abstract}

\section{Introduction}
\begin{noticebox}
    The article includes minor revisions of the paper presented at the 2024 European Signal Processing Conference (EUSIPCO), Lyon, France.
\end{noticebox}

Offline analysis of data recorded during an experiment has become the standard paradigm in neuroscience for investigating the neural underpinnings of behavior. These analyses often reveal the need to revise experimental conditions, or deficiencies in the data collection process itself, potentially rendering it uninformative about the scientific hypothesis under consideration~\cite{Paninski2018}. Consequently, experimentalists carry out multiple iterations of an experiment to obtain definitive findings, making this mode of analysis expensive and time-consuming.
The scientific questions that can be addressed from pre-collected data are also limited, especially for establishing causal links between neural activity and behavior~\cite{Panzeri2017}.
With advances in recording techniques over the past few decades, we have access to a large population of simultaneously recorded neurons during an experiment. The shift from studying single neurons to population level activity has allowed us to probe more intricate neural phenomena and has significantly enhanced our understanding of neural computation.
However, the increased spatiotemporal complexity of these recordings has made real-time analysis a challenging endeavor.

The field of \textbf{real-time neuroscience} consists of tools from statistical signal processing and machine learning that are designed to infer the underlying state of a system from streaming data samples at a \SIrange{1}{100}{\milli\second} timescale and provide a potential avenue for online analysis.
In principle, these methods can identify relevant neural states while data is being recorded, effectively closing the loop between analysis and experiments.
Such a setting naturally affords \textbf{(1) efficient data collection via adaptive experimental paradigms} and is conducive for \textbf{(2) investigating causal relationships between the neural state and behavior}. Furthermore, the ability to characterize and perturb neural states in a closed-loop setting in accordance with neuroscientifically relevant control objectives could offer significant insights into neurological disorders, and prompt the development of novel treatments.

Although existing real-time inference and control methods have demonstrated promising results, there are still conceptual and technical challenges that hinder their application in experimental settings. However, with recent developments in machine learning, specifically in sequence-to-sequence modeling and meta-learning, we are at a pivotal moment to address some of these issues. In the following sections, we present an overview of state-of-the-art approaches for identifying neural states from experimental recordings, along with neuroscientifically relevant control objectives. We highlight the challenges faced by existing approaches, and underscore the importance of large-scale, integrative efforts in Neuroscience, in tandem with further development in methods for online control and meta-learning as promising research directions for bringing real-time experimental paradigms to the forefront of Neuroscience.

\begin{figure*}[h]
    \centering
    \includegraphics[width=0.9\textwidth]{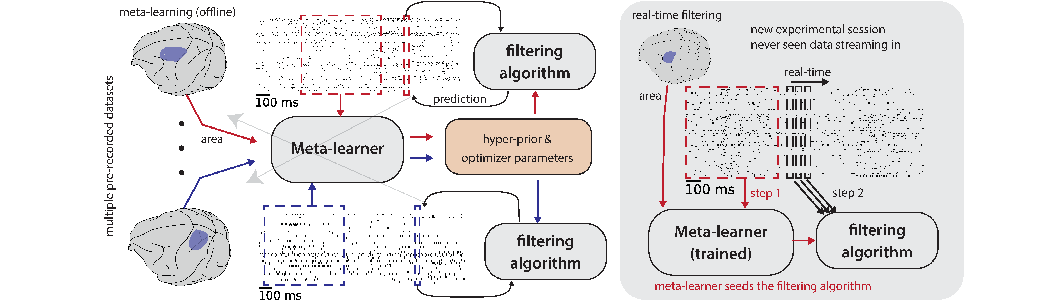}
    \caption{The meta-learner can be trained offline on pre-recorded datasets to provide inductive biases to filtering algorithms by optimizing their predictive performance (left). During an experiment, the trained meta-learner can be fine-tuned on initial samples to produce optimal model initialization, hyperparameter values, etc., and seed the online filtering algorithm (right).} 
    \label{meta_learning}
\end{figure*}

\section{State-of-the-art Approaches}\label{sec:sota}
\subsection{Non-real-time methods}
The spatiotemporal complexities of partially observed, high-dimensional neurophysiological recordings necessitates the development of methods that can extract structure from noisy data. Unsurprisingly, latent variable models (LVMs) have become ubiquitous for population level analysis of neural recordings.
However, selecting appropriate LVMs can be difficult, even for offline analysis; it is necessary to assess the amount and quality of data, the task complexity, and available computational resources.
In some cases, simple approaches such as principal components analysis or factor analysis may be suitable and are frequently used~\cite{Williams2021,Hurwitz2021} but they ignore the multi-scale dynamical structure of latent variables that can be exploited for better inference~\cite{yu_cunningham_gpfa_2009,Dowling2023c}.

State-space models (SSMs) directly specify a stochastic differential or difference equation that describes the neural state dynamics, making them capable of capturing the rich temporal structure of neural population activity~\cite{kailath2000linear,Dowling2023c}. The dynamics model can include broad inductive biases arising from neurobiology such as local smoothness, finite firing rates of neurons, oscillations, multi-stability, and so on. However, a lack of a priori knowledge of the governing law of the neural system necessitates a data-driven approach for learning model parameters~\cite{Zhao2016d}. These dynamics can be flexibly chosen -- on one end of the spectrum, linear dynamical systems can be used to model neural state evolution that afford closed-form inference and learning of the generative model parameters via expectation maximization (EM)~\cite{Paninski2009,durbin2012time,kitagawa1996monte}; on the other end, nonlinear dynamical system models might be favorable due to their high expressivity, but require approximate inference methods such as variational inference or particle filtering~\cite{doucet_nonlinear_book_2014,Zhao2019a}.

\subsection{Filtering states}
The SSM framework is particularly well suited for real-time applications since the statistical beliefs over the latent state can be recursively updated in a principled manner through recursive Bayesian estimation~\cite{sarkka2013bayesian,Chen2015}.
For a known linear dynamical system with Gaussian observations, recursive estimation of the \textit{filtering posterior} is solved through the application of the well-known Kalman filter~\cite{kalman1960new}.
However, even in the simple case of linear dynamics with unknown parameters, real-time estimation of the latent state in tandem with learning the system parameters becomes difficult.

For example, in the linear Gaussian case, given a \textit{warm-up} period before filtering estimates are accessed for data collection, subspace identification methods can be used to produce estimates of the state-space model parameters~\cite{van2012subspace}; uncertainty in parameters arising from a coarse-grained initialization during the warm-up period could be further refined through application of the Schmidt-Kalman filter~\cite{jazwinski2007stochastic}.
In practical scenarios, three immediate problems arise: (1) a nonlinear dynamical systems model may be more appropriate, (2) a long warm-up period for model parameter estimation may not be feasible, and (3) an enlarged state-space might be required to account for uncertain or poorly estimated model parameters, making the time complexity of real-time inference algorithms prohibitive\cite{jazwinski2007stochastic}.

For nonlinear dynamical system models, approximate Bayesian inference algorithms based on particle filtering or variational inference have become popular~\cite{Campbell2021-yc,Zhao2019a,Dowling2023a,crisan2018nested}. The general idea for learning the system dynamics is to recursively update an approximate posterior to the filtering posterior, then approximate the log-marginal likelihood in a differentiable manner so that the corresponding generative model parameters can be learned online. Alternatively, it is possible to tile the neural state-space and learn transitions between the tiles~\cite{Draelos2021-un}.
\begin{figure*}[h]
    \centering
    \includegraphics[width=\textwidth]{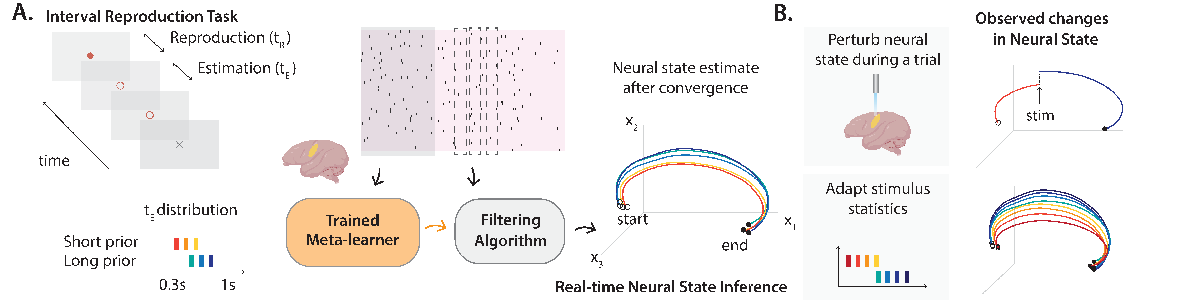}
    \caption{\textbf{A.} (Left) Example experimental setup adapted from~\cite{SOHN2019934} for an Interval Reproduction task where subjects estimate a time interval ($t_E$) and reproduce it ($t_R$). (Right) The initial recording samples are fed to the trained meta-learner to seed the online filtering algorithm and obtain the neural state estimate. \textbf{B.} In this real-time setup, the experimentalist can perform targeted neural stimulation to investigate causal links between the neural state and behavior (top), or adapt the stimulus statistics during the experiment in order to test specific hypothesis (bottom) by observing changes in the neural state.}
    \label{fig:eg_rt_perturb}
\end{figure*}
\subsection{Control objectives}
Perturbation approaches in neuroscience have been primarily used in an open-loop, for instance, in order to investigate the effect of stimulation techniques on the neural state~\cite{shea2022}, or to establish a causal role of specific brain areas in behaviors of interest\cite{Bradley2022}.
Real-time SSM opens the door for applying online control---holding the internal brain state to be constant (as done in voltage-clamping for a single neuron), and applying trajectory control to induce a sequence of neural states.
However, unlike in engineering applications, (1) we do not have a good model of the plant a priori, and (2) full control of the nervous system is ill-defined, undesirable, and perhaps entirely unfeasible.
As a result, we have previously proposed myopic control of the latent state as an alternative for neuroscientific settings~\cite{Hocker2019a}, similar to dynamic clamping in single neurons, to allow for full autonomy and agency of the larger system.
We have also proposed exploration and active sampling as control objectives for improving the sample efficiency of nonlinear system identification~\cite{etm}.

\section{Challenges}\label{sec:challenges}

\subsection{Convergence and Sample Efficiency}
As discussed in Section \ref{sec:sota}, SSMs can be readily used to infer the latent state and learn the parameters of the generative model from streaming data. However, online estimation necessitates constant space and time complexity and the optimization of these models is often non-convex and data intensive~\cite{Draelos2021-un,Zhao2019a}. This is because unlike offline training, each recorded sample is processed only once during optimization and consequently, a substantial number of samples are required for these models to attain operational efficacy. This results in undesirable experimental costs and hampers practical application, as experimental durations are typically restricted. While these issues could be partly alleviated by incorporating prior knowledge in the form of parameter initializations or optimal hyperparameter values, these models are typically trained from scratch due to a lack of inductive biases for neural dynamics.

\subsection{Identifiability}
One of the primary aims of a closed-loop setup is understanding the dynamical system underlying neural computation. The inherent nonconvex nature of the problem and utilization of optimization techniques such as stochastic gradient descent make this difficult. Moreover, in conventional SSM architectures, the latent states undergo translational and rotational equivalence throughout the joint training process~\cite{Zhao2017c}, leading to perpetual changes in the latent representation, while maintaining equivalent functionality, such as co-rotating observations and latents. This issue is unique for online learning since the past time steps are not updated accordingly with new data unlike in the offline case.

Moreover, the resultant models also lack uniqueness in their optimal parameter values, presenting a hurdle for interpretation and subsequent quantitative analyses~\cite{roeder2021linear, Durstewitz2023nn}. These models may offer mechanistic explanations for neural phenomena and produce latent manifolds that closely resemble one another but they might diverge in task-irrelevant characteristics such as velocity of the state evolution and noise. In this setting, model selection becomes a complicated process and prompts an essential inquiry: are the observed characteristics faithful reflections of the neural dynamics, or are they simply artifacts resulting from the particular model choice? This question underscores the need for thorough investigation and critical evaluation of models derived from data.

\subsection{Incorporation of knowledge}
Popular state-of-the-art probabilistic modeling approaches typically use expressive differentiable function approximators to model the underlying dynamical systems in a black-box manner~\cite{Pandarinath2018, Dowling2024b, Zhao2017c, Zhao2019a}. This modeling choices could be motivated by the trainability of these architectures, or desired properties of the learnt function.
However, there has been a notable absence of principled approaches for initializing or integrating a priori knowledge into these models, particularly concerning the characteristics of dynamical systems.
This deficiency impedes the utilization of the extensive available datasets of neural recordings and the examination of specific scientific hypotheses regarding neural dynamics without presumptions about particular functional forms.

\section{Perspectives, Opportunities, and Strategies}
\subsection{Large-scale Integrative Neuroscience}
Neural systems are remarkably intricate, both in terms of function and anatomy. Due to their inherent complexity, experimental paradigms and analysis in Systems neuroscience are often designed to probe the function of specific areas of the brain.
Whereas this has offered valuable insights into distinct regions of the brain, it does not shed light on how interaction between various parts of the nervous system give rise to behavior.
We argue that a model-centric integration process for consolidating existing knowledge and unifying computational models of neural recordings is essential for making significant progress in this direction. This would involve substantial efforts to standardize neural data and model fitting, as well as higher quality control for algorithms and their implementations, necessitating large-scale collaboration across research groups. Such a process would also be essential for Embodied AI settings, where the integrated model can be deployed as an interactive system and used as a testbed for complex real-time experiments~\cite{Zador2023}. 

Apart from improving scientific understanding, this integrated model can be useful for constraining the search space of online filtering algorithms, thereby reducing the problem complexity. For instance, it is widely known that the neural dynamics underlying neural activity in motor cortex across animals engaged in similar tasks show qualitative similarities as well~\cite{safaie_2023}. This has been leveraged to transfer knowledge in the form of expected latent dynamics across recordings~\cite{Vermani2023b}. Large-scale integrative models can provide further inductive biases for diverse settings and enable faster convergence of online filtering algorithms. These inductive biases can complement biologically informed constraints and parameterizations of latent dynamics to improve identifiability of learnt solutions~\cite{sslds2024,schimel2024}.
\subsection{Meta Learning}
Meta-learning approaches in machine learning aim to train models that generalize to novel settings with few training samples~\cite{ren18fewshotssl, pmlr-v70-long17a}. However, there has been limited work on applying these methods for improving the convergence speed and sample efficiency of online filtering methods. 

There are two broad categories of meta-learning approaches -- gradient-based methods such as model agnostic meta-learning learn parameter initializations for faster convergence on downstream tasks~\cite{pmlr-v70-finn17a}; in contrast, model-based approaches train a black-box meta-learner along with the base model to directly generate parameters of the base model~\cite{edwards2017towards, pmlr-v70-munkhdalai17a}. These methods can be applied to meta-learn parameter initializations, optimal learning rate, latent state dimensionality, etc., by optimizing the performance of the SSM on diverse, pre-collected recordings offline. This trained meta-learner can subsequently be fine-tuned at the beginning of the experiment to provide data-driven priors to online filtering algorithms (Fig.~\ref{meta_learning}).
While this offers a promising solution for improving model convergence and sample efficiency, existing meta-learning approaches need to be adapted to deal with statistical heterogeneities across different neural datasets, for instance, differences in the number and tuning properties of recorded neurons, the dimensionality of neural state, and to incorporate prior knowledge  and structural hypotheses.

\subsection{Novel Control Methods}
Despite advances in recording technologies, a concomitant increase in the number of model parameters for analyzing this data has paradoxically diminished the statistical power to answer scientific questions. The problem is magnified by inefficient sampling of high-dimensional neural dynamics. Closed-loop active learning is a promising direction for facilitating data-efficient nonlinear system identification~\cite{Chen2021}, where the algorithm determines the most informative data for the model inference based on the current belief via Bayesian experimental design.
In traditional closed-loop stimulation, the cost function is defined using explicitly observable neural recordings and responses~\cite{bolus2018}. Real-time SSM would enable closed-loop active learning with cost functions defined over the latent state, allowing the model to fully utilize rich temporal structure such as attractor dynamics, while making the control objectives more interpretable and robust to noise.
\begin{figure}[t!h]
    \vspace{-5pt}
    \centering
    \includegraphics[width=0.8\linewidth]{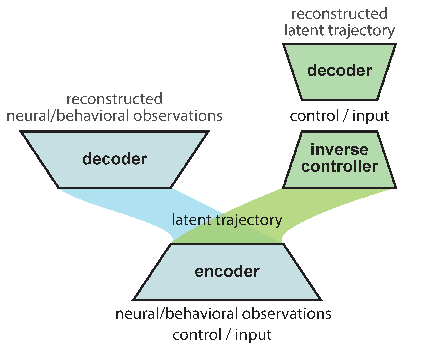}
    \caption{
	Using two seq-to-seq autoencoders, first one to infer, impute, and forecast a low-dimensional neural trajectory, and the second one to act as an inverse controller to map desired neural trajectory to control signal.
    }
    \label{fig:seq2seqAE}
\end{figure}

In addition to enabling system identification, the development of control objectives is imperative for investigating complex scientific phenomena.
Conventionally, model predictive control (MPC) employs the current state, observations and external inputs to predict the system's next state. 
A longer prediction horizon is achieved by recursively feeding the model's output back as the new input, along with the remaining variables, in a {\em rolling horizon forecast} fashion. 
This approach, while straightforward in theory, is both costly, as it requires the entire model to perform multiple sequential inferences, and susceptible to accruing inaccuracies over time, as it will propagate its own prediction error.
We foresee an opportunity in MPC to leverage the advances in differentiable and computationally efficient sequence-to-sequence modeling that revolutionized language models.
For example, causal transformers could infer latent processes\cite{Jaegle2021-rb} and serve as inverse controllers in an autoencoder framework (Fig.~\ref{fig:seq2seqAE}).
The inference and control problems can be parallelized and the latency can be reduced via scaling computing hardware, enabling better real-time control through faster generation of candidate sequences for neural perturbation.

\section{Discussion}
There are still several engineering challenges that need to be solved to achieve a paradigm shift from offline analysis to short-latency closed-loop experiments in Neuroscience. 
The next generation of neurotechnology will push the boundaries of fundamental Neuroscience by enabling more complex experiments (Fig. \ref{fig:eg_rt_perturb}). It could also have significant impact on our understanding of various neurological disorders and provide novel treatment options through innovations in clinical devices. Along with technical developments, collaborative efforts will be imperative for solving major hurdles. 
We advocate for large-scale, collaborative initiatives that promote reproducibility, data standardization and open science in order to fully realize this goal and advance the neuroscientific community at large.

\bibliography{ref,catniplab}
\bibliographystyle{IEEEtran_memming_v1}

\begin{IEEEkeywords}
real-time neuroscience, closed-loop control, state-space model, nonlinear dynamical system
\end{IEEEkeywords}

\end{document}